\documentclass{article}

\usepackage[preprint]{neurips_2025}

\usepackage[utf8]{inputenc} 
\usepackage[T1]{fontenc}    
\usepackage{hyperref}       
\usepackage{url}            
\usepackage{booktabs}       
\usepackage{amsfonts}       
\usepackage{nicefrac}       
\usepackage{microtype}      
\usepackage{wrapfig}
\usepackage[pdftex]{graphicx}
\usepackage[table]{xcolor}
\usepackage{subcaption}

\newcommand{\model}{Nequix}
\definecolor{oursrow}{HTML}{E3F2FD} 

\title{Training a Foundation Model for Materials on a Budget}

\author{
    Teddy Koker\qquad Mit Kotak\qquad Tess Smidt\\
    Department of Electrical Engineering  and Computer Science\\
    Massachusetts Institute of Technology\\
    Cambridge, MA 02139 \\
    \texttt{\{tekoker,mkotak,tsmidt\}@mit.edu} \\
}

\begin{document}

\maketitle

\begin{abstract}
  Foundation models for materials modeling are advancing quickly, but their
  training remains expensive, often placing state-of-the-art methods out of
  reach for many research groups. We introduce \model{}, a compact
  E(3)-equivariant potential that pairs a simplified NequIP design with modern
  training practices, including equivariant root-mean-square layer normalization
  and the Muon optimizer, to retain accuracy while substantially reducing
  compute requirements.  \model{} has 700K parameters and was trained in 100
  A100 GPU-hours. On the Matbench-Discovery and MDR Phonon benchmarks, \model{}
  ranks third overall while requiring a 20 times lower training cost than most
  other methods, and it delivers two orders of magnitude faster inference speed
  than the current top-ranked model. We release model weights and fully
  reproducible codebase at \url{https://github.com/atomicarchitects/nequix}.
  
\end{abstract}

\section{Introduction}

Machine learned inter-atomic potentials (MLIPs) are rapidly improving in
capability and scope, with foundation models trained on broad datasets of
atomistic materials offering the promise of augmenting or replacing expensive
\textit{ab initio} density functional theory (DFT) calculations
\citep{batatia2023foundation}. While performance on community benchmarks such as
Matbench-Discovery \citep{riebesell2025framework} is rising, the computational
costs of both data generation and curation as well as the training of MLIP
models on these datasets remain prohibitively expensive for many labs.

We pursue an orthogonal goal to scaling: a lower computational cost recipe that
preserves strong downstream accuracy. Concretely, we revisit a simplified
E(3)-equivariant architecture based on NequIP \citep{batzner20223} with modern training practices: root-mean-square layer normalization for stability, \mit{latest custom CUDA kernels \citep{openequivariance}}, and optimizer choices inspired by
``speedrunning'' deep learning workflows \citep{modded_nanogpt_2024}. The
resulting model, \model{}, has 700K parameters and can be trained in \mit{100} GPU
hours, while remaining competitive with larger and more costly to train models
on Matbench-Discovery and other phonon prediction tasks.

Our contributions are threefold: (1) a simplified NequIP architecture featuring
an equivariant layer normalization and efficient JAX and PyTorch implementations; (2) a
budget-conscious training pipeline leveraging the Muon optimizer
\citep{jordan2024muon}, achieving fast convergence; and (3) evaluations on the
Matbench-Discovery and MDR phonon \citep{loew2025universal} benchmarks. Compared
to prior MPtrj-trained models
\citep{chen2022universal,deng2023chgnet,batatia2023foundation,bochkarev2024graph,neumann2024orb,barroso2024open,fu2025learning,zhang2025graph,yan2025materials},
we rank \textit{third} (as of August 2025) on both benchmarks at 1/20 the
training cost of any other published model and with 100$\times$ faster inference
than the current top-ranking model.

\section{Methods}
\label{methods}

\begin{figure}
  \centering
  \makebox[\linewidth][c]{%
    \subcaptionbox{}{\includegraphics[height=0.259\linewidth]{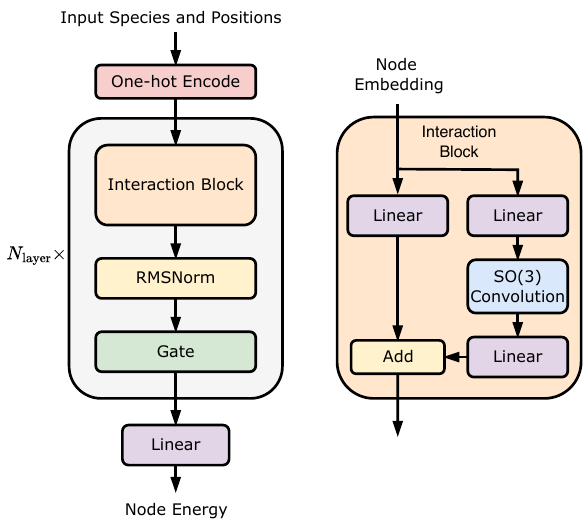}}\hspace{0.02\linewidth}%
    \subcaptionbox{}{\includegraphics[height=0.259\linewidth]{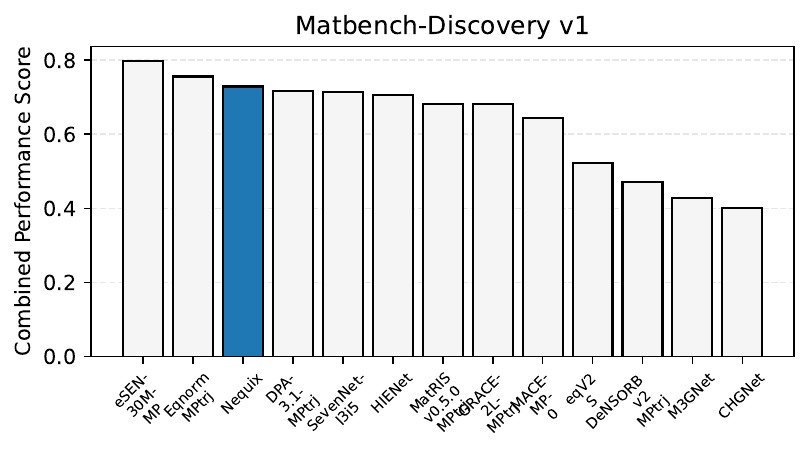}}\hspace{0.01\linewidth}%
    \subcaptionbox{}{\includegraphics[height=0.259\linewidth]{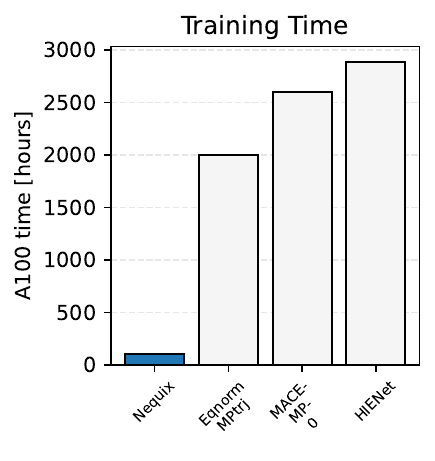}}%
  }
  \caption{(a) \model{} architecture, a simplified version of NequIP
  \citep{batzner20223}, with a species-independent residual connection and layer
  normalization. (b) Combined performance scores of compliant models on the
  Matbench-Discovery (unique prototypes subset), collected on 2025-08-17. (c) Available published training times
  of current compliant models.} 
  \label{fig:one}
\end{figure}

\paragraph{Architecture} \model{} follows a simplified version of the NequIP
\citep{batzner20223} architecture, as shown in figure \ref{fig:one}a. We adopt
two modifications suggested by \cite{park2024scalable}: the species-specific
self-connection layer within the interaction block is replaced with a single
linear layer, and unused non-scalar representations are discarded from the final
layer. Lastly, we add an equivariant root-mean-square layer normalization
(RMSNorm) \cite{liao2023equiformerv2}, which we find improves performance in our
optimization setting. We document the full architecture hyper-parameters and the
rationale behind each decision in Table \ref{tab:hparams}.

\paragraph{Implementation} \model{} is implemented in both JAX
\citep{jax2018github, kidger2021equinox} and PyTorch\citep{pytorch, pytorch2},
taking advantage of just-in-time compilation and efficient automatic
differentiation. Following standard energy-conserving MLIP practice
\citep{fu2025learning}, forces are obtained as the negative energy gradient with
respect to atomic positions, $-\nabla_{\mathbf r} E$, and stresses as the energy
derivative with respect to strain, normalized by volume, $\sigma =
V^{-1}\,\partial E/\partial \varepsilon$, where $E$ is the predicted total
energy of the system, $\mathbf r$ is an atom position, $\varepsilon$ is the
strain tensor, $V$ is the simulation cell volume, and $\sigma$ denotes the
stress tensor.

\paragraph{Dynamic batching} Materials radius graphs vary widely in their
numbers of nodes and edges due to differences in unit-cell size and atom count.
With fixed-size batching, the largest graphs dictate memory usage, leaving GPUs
underutilized for most batches. To keep batch workloads more uniform while
respecting memory limits, we use dynamic batching \citep{speckhard2025analysis}:
each batch is filled up to caps on the total nodes and edges. We set these caps
to $1.1 \times$ (ideal batch size) $\times$ (dataset-average nodes-per-graph and
edges-per-graph), respectively. Batches are then padded to these caps to fulfill
the static shape requirements of JAX \citep{jraph2020github}. This capability is
only used for the JAX implementation.

\begin{figure}
  \centering
  \includegraphics[width=\linewidth]{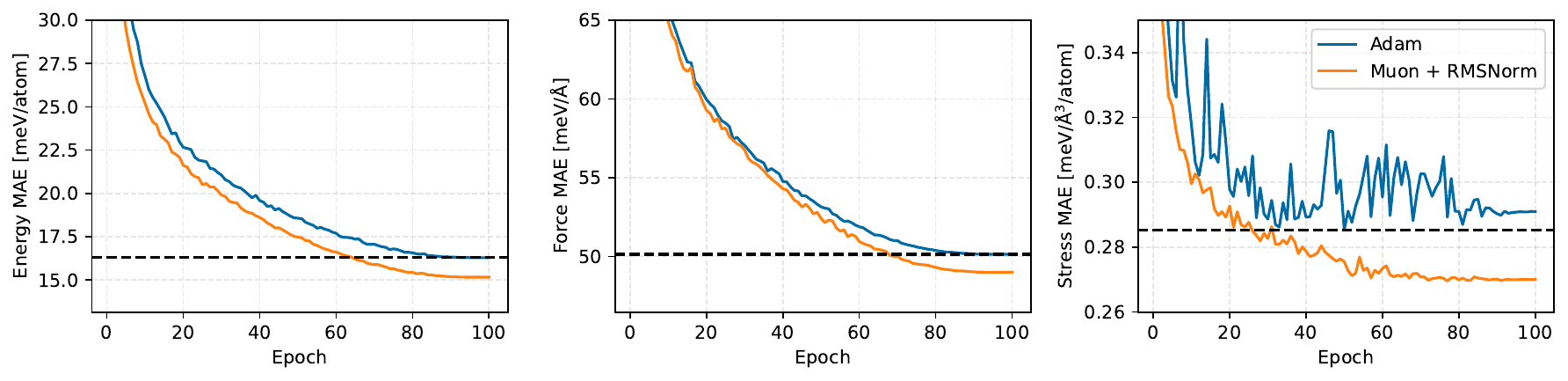}
  \caption{Validation metrics during training of a smaller version of \model{} configuration
  with Adam and Muon, trying learning rates in $\{0.03,0.01,0.003,0.001\}$ and
  with/without RMSNorm. This model configuration uses the same hyperparameters
  as the final model, except with hidden irreps of \texttt{128x0e\,+\,64x1o}.
  The dotted horizontal line shows the best validation performance reached
  during the Adam training.}
  \label{fig:muon} 
\end{figure}

\paragraph{Optimization and normalization} We compare the widely used Adam
optimizer \cite{kingma2014adam} with the recently proposed Muon optimizer
\cite{jordan2024muon}, which uses the Newton-Schulz algorithm to orthogonalize
weight updates. Using a smaller version of \model{} on MPtrj
\citep{jain2013commentary, deng2023chgnet} with hidden irreps of
\texttt{128x0e\,+\,64x1o}, we sweep learning rates of
$\{0.03,0.01,0.003,0.001\}$ for Adam and Muon, each with and without RMSNorm.
The lowest validation error runs for each optimizer are shown in Fig.
\ref{fig:muon}. We find that the Muon configuration achieves comparable
energy/force errors to Adam in 60-70\% of the epochs, and results in a $7\%$
reduction in energy MAE. We also find a significant reduction in the variance of
stress error, which we notice in runs that use RMSNorm. Notably, the presence of
the RMSNorm layer generally resulted in lower validation error for Muon-based
training configurations, and higher for those using Adam.

\paragraph{GPU kernels} There has been recent work \citep{openequivariance,
cuequivariance, newquip, flashtp} on writing custom GPU kernels for the
expensive equivariant tensor product operation \citep{priceoffreedom}. These
kernels fuse the tensor product and the outer gather-scatter from the message
passing step into a single GPU kernel. This avoids storing costly edge-based
intermediates in GPU memory, improving both runtime and memory usage.

\paragraph{Training procedure} The final \model{} model is trained for 100
epochs on MPtrj \citep{jain2013commentary, deng2023chgnet}, of which we hold out
5\% for validation. More details on the training settings are provided in Sec.
\ref{configuration}. The model was trained on 2 NVIDIA A100 80 GB GPUs in 50 hours,
for a total cost of 100 GPU hours.

\section{Experiments}

\subsection{Matbench-Discovery benchmark}

Matbench-Discovery \citep{riebesell2025framework} provides a standard framework
for evaluating interatomic potentials in a high-throughput materials screening
task consisting of geometry optimization and energy prediction on a set of
257,487 generated structures, and thermal conductivity prediction on a set of
103 structures. Ground truth is calculated with DFT/PBE level of theory, the
same as MPtrj. The primary metrics include: 1) the F1 for stable/unstable
classification after relaxation; 2) root mean squared displacement (RMSD)
between predicted and reference structures after relaxation; and 3) symmetric
relative mean error in predicted phonon mode contributions to thermal
conductivity $\kappa$ ($\kappa_\mathrm{SRME}$). A normalized and weighted
combination of these metrics are then used to compute a combined performance
score (CPS-1), which is used for ranking.

Following \cite{riebesell2025framework}, we integrate our interatomic potential
as Atomic Simulation Environment (ASE) calculator, which is then used to perform
structure relaxation and phonon calculations with the default settings of the
benchmark. For comparison, we consider only models in the compliant subset of
the benchmark. This consists only of models that are trained on MPtrj or
subsets, which limits data leakage and offers a more fair comparison among
methods. Table \ref{tab:matbench} contains the performance of \model{} along
with all current compliant models at the time of writing. We also include the
reported training cost for the models when available, visualized in Fig.
\ref{fig:one}. We find that \model{} ranks third by CPS-1, outperforming most
models at a fraction of the training cost. It is noteworthy that this high
ranking is due to high performance in the thermal conductivity task, however, the
F1 score is still comparable to many of the other methods.

\begin{table}[htbp]
  \centering
  \small
  \caption{Matbench-Discovery v1 compliant leaderboard, sorted by combined
  performance score (CPS-1). Metrics are shown for the unique prototypes subset.
  Train cost is measured in A100 hours. Data as of 2025-08-17. } \label{tab:matbench} 
  \begin{tabular}{lcccccc}
      \toprule
      Model & Params & \shortstack{Train cost}  & RMSD$\downarrow$   & $\kappa_{\mathrm{SRME}}\downarrow$ & F1$\uparrow$  & CPS-1$\uparrow$ \\
      \midrule
      eSEN-30M-MP                & 30.1M & -     & 0.075 & 0.340 & 0.831 & 0.797 \\
      Eqnorm MPtrj               & 1.31M & 2000  & 0.084 & 0.408 & 0.786 & 0.756 \\
      \rowcolor{oursrow}\model{} & 708K  & 100   & 0.085 & 0.446 & 0.750 & 0.729 \\
      DPA-3.1-MPtrj              & 4.81M & -     & 0.080 & 0.650 & 0.803 & 0.718 \\
      SevenNet-l3i5              & 1.17M & -     & 0.085 & 0.550 & 0.760 & 0.714 \\
      HIENet                     & 7.51M & 2888  & 0.080 & 0.642 & 0.777 & 0.707 \\
      MatRIS v0.5.0 MPtrj        & 5.83M & -     & 0.077 & 0.861 & 0.809 & 0.681 \\
      GRACE-2L-MPtrj             & 15.3M & -     & 0.090 & 0.525 & 0.691 & 0.681 \\
      MACE-MP-0                  & 4.69M & 2600  & 0.092 & 0.647 & 0.669 & 0.644 \\
      eqV2 S DeNS                & 31.2M & -     & 0.076 & 1.676 & 0.815 & 0.522 \\
      ORB v2 MPtrj               & 25.2M & -     & 0.101 & 1.725 & 0.765 & 0.470 \\
      M3GNet                     & 228K  & -     & 0.112 & 1.412 & 0.569 & 0.428 \\
      CHGNet                     & 413K  & -     & 0.095 & 1.717 & 0.613 & 0.400 \\

      \bottomrule
  \end{tabular}
\end{table}

\subsection{MDR phonon benchmark}

Performance is also evaluated on the MDR phonon benchmark
\citep{loew2025universal}, a set of 10,000 phonon calculations also done with
DFT/PBE level of theory. We follow the identical procedure to
\citet{loew2025universal}, first performing a geometry relaxation, then phonon
calculations using displacements of supercells. We report the mean absolute error (MAE) of
properties derived from the phonon calculation: maximum phonon frequency
$\omega_{\max}$, vibrational entropy $S$, Helmholtz free energy $F$, and heat
capacity at constant volume $C_V$. Table \ref{tab:mdr_phonon} demonstrates the
performance of \model{} compared to other MPtrj-trained models. Similarly to
Matbench-Discovery, we achieve performance within the top three of models, with
a fraction of the parameter count of other methods.

\begin{table}[htbp]
  \centering
  \small
  \caption{Model performance of MPtrj-trained models on the MDR phonon
  benchmark, sourced from \citet{loew2025universal} and \citet{fu2025learning}. Metrics are MAE of maximum phonon frequency $\omega_{\max}$ (K), vibrational entropy $S$ (J/K/mol), Helmholtz free energy $F$ (kJ/mol) and heat capacity at constant volume $C_V$ (J/K/mol).}
  \label{tab:mdr_phonon}
  \begin{tabular}{lcccc}
      \toprule
      Model & MAE($\omega_{\max}$) & MAE($S$) & MAE($F$) & MAE($C_V$) \\
      \midrule
      eSEN-30M      & 21 & 13 & 5 & 4 \\
      SevenNet-l3i5 & 26 &  28 & 10 &  5 \\
      \rowcolor{oursrow}\model{} & 26 & 33 & 12 & 6 \\
      SevenNet-0    & 40 &  48 & 19 &  9 \\
      GRACE-2L (r6) & 40 &  25 &  9 &  5 \\
      MACE           & 61 &  60 & 24 & 13 \\
      CHGNet         & 89 & 114 & 45 & 21 \\
      M3GNet         & 98 & 150 & 56 & 22 \\
      \bottomrule
  \end{tabular}
\end{table}

\subsection{Inference speed}

\begin{wrapfigure}{r}{0.4\textwidth}
  \vspace{-1.7em}
  \centering
  \includegraphics[width=\linewidth]{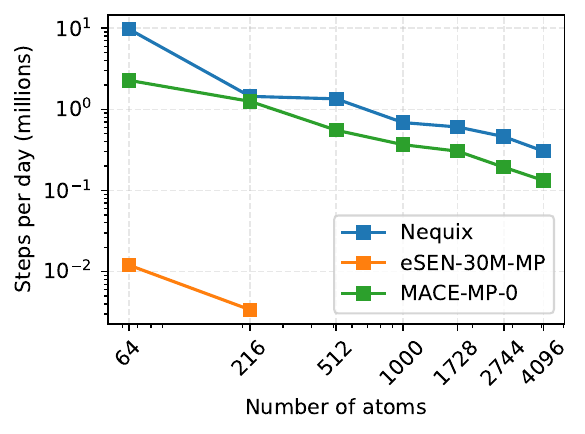}
  \caption{Inference speed of various models in steps per day.}
  \label{fig:inference}
  \vspace{-2em}
\end{wrapfigure}

Finally, we compare inference speed with existing interatomic potentials by
using ASE \cite{larsen2017atomic} to run calculations on diamond across varying
unit cell sizes, timing the calculations for each model.\footnote{See
\url{https://github.com/mitkotak/matbench-speed} for inference speed
benchmarking code.}. We run each model in the default configuration in which it
used within its ASE calculator, with compilation and kernels wherever specified
in the documentation. See Section \ref{inference} for more information on benchmarking setup.


Figure \ref{fig:inference}
compares performance of each model in terms of steps per day vs. number of
atoms. In this study, \model{} is about 100$\times$ faster than eSEN, offering a new option in the accuracy vs. speed
Pareto frontier at a fraction of the training cost.

\section{Conclusion}

We presented \model{}, an E(3)-equivariant interatomic potential that pairs a
simplified NequIP architecture with modern training practices. Our results show
that \model{} achieves competitive accuracy on Matbench-Discovery and the MDR
phonon benchmark at less than one quarter of the reported training cost of many
contemporaries. This resource-efficient recipe provides a practical alternative
to large-scale foundation models and helps broaden access to high-quality
atomistic modeling in settings with more limited compute. We release trained
weights and a JAX/PyTorch codebase to streamline reuse and extension.

Looking ahead, we see several promising directions: scaling training duration
and data while maintaining budget discipline, exploring pretraining and
fine-tuning regimes across broader datasets, and pushing cost even lower through model
distillation, pruning, quantization, kernel implementations, or more
data-efficient training. We hope \model{} serves as a strong, efficient baseline
for future work on accessible materials foundation models.

\begin{ack}

This work was supported by the National Science Foundation under Cooperative
Agreement PHY-2019786 (The NSF AI Institute for Artificial Intelligence and
Fundamental Interactions, \url{http://iaifi.org/}) NSF Graduate Research Fellowship program under Grant No. DGE-1745302, and by DOE ICDI grant DE-SC0022215.
This research used resources of the National Energy Research Scientific
Computing Center (NERSC), a Department of Energy User Facility using NERSC award
ERCAP0033254.

\end{ack}

{
\small

\bibliographystyle{unsrtnat}
\bibliography{refs}

}


\appendix
\setcounter{table}{0}
\setcounter{figure}{0}
\renewcommand{\thetable}{A.\arabic{table}}
\renewcommand{\thefigure}{A.\arabic{figure}}

\section{Appendix}

\subsection{Training and model configuration}
\label{configuration}

\begin{figure}[htbp]
  \centering
  \includegraphics[width=\linewidth]{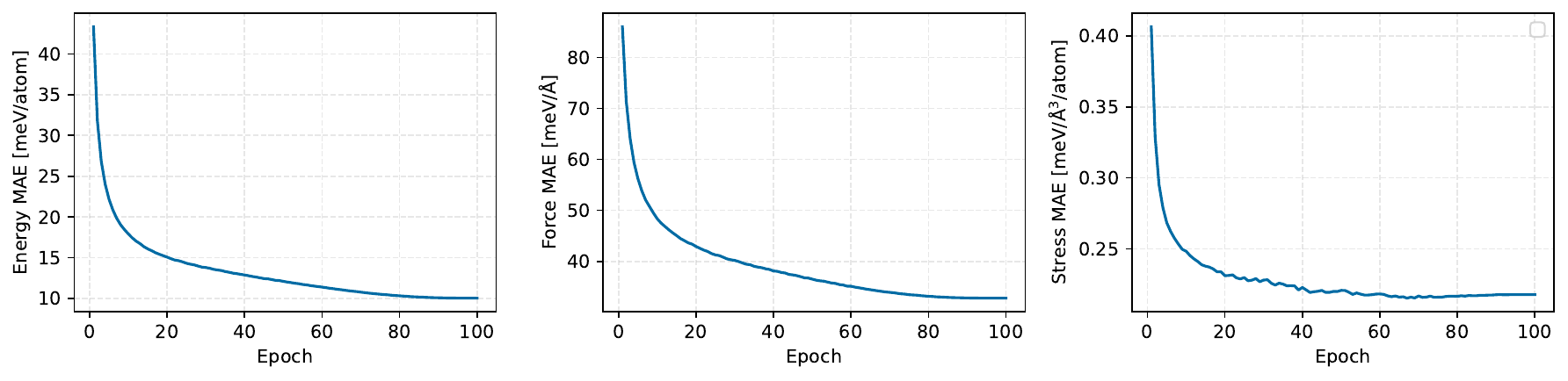}
  \caption{Validation curves for Nequix training on MPtrj.}
  \label{fig:final_training} 
\end{figure}

Table \ref{tab:hparams} shows the hyper-parameters used to train \model{}. The
model is trained for 100 epochs, using an MAE loss function on energy and
stress, and $l_2$ loss on forces. We use a linear warmup with cosine decay
learning rate schedule. Figure \ref{fig:final_training} shows the energy, force,
and stress MAE on the validation set throughout training. The final MAEs are
10.05 meV/atom, 32.79 meV/\AA, and 0.22 meV/\AA$^3$/atom for energy, forces,
and stress respectively.

\begin{table}[htbp]
    \centering
    \small
    \caption{Hyper-parameters used and rationale behind selection}
    \label{tab:hparams}
    \begin{tabular}{lcp{6cm}}
        \toprule
        Hyper-parameter & Value & Notes/Rationale \\
        \midrule
        Radial cutoff & 6 \AA & Most models use 5 or 6 \AA; 6 performed slightly better in preliminary validation performance. \\
        Hidden irreps & \texttt{128x0e\,+\,64x1o\,+\,32x2e\,+\,32x3o} & From SevenNet-l3i5. \\
        $L_\mathrm{max}$ & 3 & Consistent with hidden irreps. \\
        $N_\mathrm{layers}$ & 4 & Balance of performance and efficiency. \\
        Radial basis size & 8 & From NequIP and analysis from Sec. 5.2 of~\cite{fu2025learning} \\
        Radial MLP size & 64 &  From NequIP. \\
        Radial MLP layers & 2 & From NequIP. \\
        Polynomial cutoff $p$ & 6.0 & From NequIP. \\
        Radial basis function & Bessel & From NequIP. Also tried Gaussian, which had minimal difference on validation performance.\\ 
        Learning rate & 0.01 & Selected from $\{0.03, 0.01, 0.003, 0.001\}$ based on validation performance early in training. \\
        Warmup epochs & 0.1 & From eSEN. \\
        Warmup factor & 0.2 & From eSEN. \\
        Optimizer & Muon & See Sec. \ref{methods}. \\
        Weight decay & 0.001 & From eSEN. Also tried 0.0, which led to worse validation performance.\\
        Energy weight & 20 & From eSEN.\\
        Force weight & 20 & From eSEN.\\
        Stress weight & 5 & From eSEN.\\
        Batch size & 256 (dynamic) & See Sec. \ref{methods} \\
        Number of epochs & 100 & Standard training duration.\\
        \bottomrule
    \end{tabular}
\end{table}

\newpage
\subsection{Inference Configuration}
\label{inference}

We summarize all of the inference configurations in \autoref{tab:ase_choices}.
We use MPTrj versions for all of the models except NequIP where the MPtrj model
did not have kernel support, so we instead use the OMat24 model after confirming
that they have the same hyperparams. For compatibility with eSEN, we use the older 
\texttt{fairchem} OCP calculator, and expect the latest version to have less overhead.

\begin{table}[htbp]
  \centering
  \caption{Inference configuration for the ASE benchmarking setup}
  \label{tab:ase_choices}
  \begin{tabular}{lcc}
       \toprule
       Model & Compile & Kernels \\
       \midrule
       eSEN-30M      & $\times$ & $\times$ \\
       MACE-MP0           & \texttt{torch.compile} & cuEquivariance \\
       Nequix         & \texttt{torch.compile} & openEquivariance \\
  \end{tabular}
\end{table}


\newpage

\end{document}